\documentstyle[11pt]{article}

\begin{document}
\title{Protogalactic Extension of the Parker Bound}
\author{M. J. Lewis\thanks{Electronic mail: {\tt
lewism@umich.edu}}, Katherine Freese\thanks{Electronic mail: {\tt
ktfreese@umich.edu}}, and Gregory Tarle\thanks 
{Electronic mail: {\tt tarle@umalp1.physics.lsa.umich.edu}} \\
Randall Physics Laboratory \\ 
University of Michigan \\
Ann Arbor, MI 48109-1120 
}
\date{\today}
\maketitle

\begin{abstract} 

We extend the Parker bound on the galactic flux $\cal F$ of magnetic
monopoles.  By requiring that a small initial seed field must survive the
collapse of the protogalaxy, before any regenerative dynamo effects become
significant, we develop a stronger bound.  The survival and continued
growth of an initial galactic seed field $\leq 10^{-9}$G
demand that ${\cal F} \leq 5
\times 10^{-21} (m/10^{17} \mbox{ GeV}) \mbox{ cm}^{-2} \mbox{sec}^{-1}
\mbox{sr}^{-1}$. For a given monopole mass,
this bound is four and a half orders of magnitude
more stringent than the previous `extended Parker bound' \cite{adams},
but is more speculative as it 
depends on assumptions about the 
behavior of magnetic fields during protogalactic collapse.
For monopoles which do not overclose 
the Universe ($\Omega_m <1$), the maximum flux allowed
is now $8 \times 10^{-19}$ cm$^{-2}$ s$^{-1}$ sr$^{-1}$, a factor
of 150 lower than the maximum flux allowed by the extended Parker bound.

\end{abstract}


\section{Introduction} 

The existence of magnetic monopoles has long been an intriguing
prospect, motivating both theorists and experimentalists for more than
50 years.  Dirac \cite{dirac} first showed
that magnetic monopoles could
be accomodated within electromagnetic theory if their
magnetic charge, $g$, is given by an
integer multiple of $\hbar c/2 e$.  
In 1974, t'Hooft \cite{tooft} and Polyakov
\cite{polyakov} independently demonstrated that monopoles are necessary
components of any Grand Unified theory (GUT) that includes
electromagnetism.  If GUTs are shown to be correct, monopoles with mass
$m \sim m_{GUT}/\alpha_{GUT} \sim 10^{15} - 10^{19} \mbox{ GeV}$ 
are expected.  However, no magnetic
monopole has ever been observed, and the cosmic abundance of monopoles
remains an open question.

The experimental search for magnetic monopoles
has intensified in the past decade.  The MACRO experiment
\cite{macro}, now fully operational, has placed stringent
limits (90\% confidence level) on the terrestrial 
monopole flux.  For a combined direct magnetic monopole search
using scintillators, streamer tubes, and nuclear track detectors,
MACRO has reached flux sensitivities below
3.2 $\times 10^{-16}$ cm$^{-2}$ sec$^{-1}$ sr$^{-1}$
for monopoles in the velocity range $10^{-4} < \beta
< 10^{-1}$.

Estimates of monopole abundance based on GUT phase transitions yield a
monopole density which overcloses the universe by several orders of
magnitude \cite{kibble}.  Inflationary models \cite{guth} can resolve this
``monopole problem'' by reducing the monopole abundance within the
observable universe to an exponentially small value.  It is difficult to
subsequently predict theoretical expectations for the number of monopoles
in the universe. Hence astrophysics can provide useful benchmark
abundances for experimenters looking for monopoles, and the most practical
constraints on the present day magnetic monopole flux arise from these
astrophysical bounds, which fall into three categories: 1) cosmological
bounds, which require that monopoles can provide at most the critical
density of the universe; 2) nucleon decay catalysis bounds, arising from
the hypothesis that monopoles catalyze nucleon decay in the cores of
neutron stars and white dwarfs;  and, 3) Parker-type bounds, which demand
that monopoles not drain energy from astrophysical magnetic fields faster
than they are regenerated.  Catalysis bounds are by far the most stringent
\cite{lots}, but depend on the value of the cross-section for monopole
catalyzed nucleon decay.  In this paper, we reconsider the Parker-type
bounds, noting that a stronger limit may be derived by considering the
survival of a magnetic seed field during the collapse of the protogalaxy. 

Parker first emphasized that the existence of observable Galactic
magnetic fields must place an upper limit on the magnetic monopole flux
\cite{parker}. The presence of a
Galactic magnetic field indicates a relative dearth of magnetic
monopoles.

The original flux bound based upon the survival of today's galactic
magnetic field is the most straightforward \cite{parker}.  Magnetic
monopoles which move along field lines absorb kinetic energy at the
expense of the field.  In order for the galactic field to survive, the
magnetic monopoles must not drain field energy faster than it is
regenerated by the dynamo, presumed to act on the order of the Galactic
rotation time period, $10^{8}$ yrs \cite{parker2}.  This implies
an upper bound on the flux of monopoles, ${\cal F} \leq
10^{-16} \mbox{ cm}^{-2} \mbox{sec}^{-1} \mbox{sr}^{-1} $. 
This bound was later shown to be mass dependent \cite{turner}.
Subsequently Adams {\it et al} \cite{adams} obtained an ``extended
Parker bound'' by requiring survival and growth of a small galactic
seed field after the collapse of the protogalaxy:
\begin{equation}
\label{extended}
{\cal F} \leq 10^{-16} (m/10^{17} {\rm GeV}) {\rm cm}^{-2} {\rm s}^{-1}
{\rm sr}^{-1} \, .
\end{equation}

Here we extend these bounds further by considering an even
earlier epoch in the history of the Galaxy, namely during
collapse of the protogalaxy to the size that it has today.
The flux bounds that we obtain in this way are the most
stringent Parker type bounds to date for three reasons:
(i) the magnetic field at this early time
is very small with a larger coherence length than today, 
(ii) the protogalaxy starts to collapse at
a higher redshift $z \sim 5$, and (iii) at this early time
no dynamo can yet be important, so that we take the primary enhancement
mechanism for the B field to be flux freezing,
${d \over dt}(BR^2) = 0$. Section 2 presents the
protogalactic extension of the Parker bound. Section 3
concludes with a discussion of caveats to this bound,
in particular the suggestion by Kulsrud $\it{et \, al}$ \cite{kulsrud}
that Kolmogorov turbulence amplifies the magnetic field.
The farther back one goes in the history
of the Galaxy, the more speculative the
bounds become, since we do not really understand the origin
of the galactic magnetic field.

\section{The Protogalactic Extension}

The origin of the Galactic magnetic field remains an outstanding
problem of theoretical astrophysics.  Faraday rotation measurements of
polarized pulsar radio emissions indicate a mean Galactic magnetic field
strength of $(2-3)\times 10^{-6}$ G.  The polarization of starlight by
aligned interstellar dust grains suggests a large scale field, coherent
over scales of (1-2) kpc, and extending more or less in the azimuthal
direction in the disk of the Galaxy.  Such fields may have originated from
a large ({\em i.e.} $\sim 10^{-9}$ G) primordial seed field amplified by
the collapse of the Galaxy, or by a much smaller seed field ($\sim
10^{-20}$ G)  that was subsequently amplified by a fast dynamo.  Recent
work by Enqvist \cite{enqvist} has argued that
a large scale $10^{-20}$ G seed field can arise via turbulent evolution from
microscopic primordial magnetic fields such as those
arising during cosmological phase transitions. Additional schemes for
obtaining seed fields by considering magnetic flux entrained in 
the winds from young
stellar objects \cite{vishniac}, or by various battery mechanisms
\cite{biermann} also exist, and under the most favorable conditions can
produce seed fields of $10^{-11}$ G. 
As an alternative to dynamo amplification of a small seed field,
Kulsrud {\em et. al.} \cite{kulsrud} argue for a protogalactic origin for
the magnetic field, wherein primary amplification of the field comes from
Kolmogorov turbulence; we will further consider this
possibility below.  The protogalactic
extension to the flux bound is based on the fact that, regardless of which
scenario is correct, the existence of a field today requires that {\em
some} field must exist after the collapse of the protogalaxy.
To be conservative, we will take the largest possible protogalactic
field value, $B_0 = 10^{-9}$G, which assumes that flux freezing
was the only early amplification mechanism.  This value is conservative
in that it will give the least restrictive bound on the monopole
flux.

Clearly the Galactic magnetic field is most vulnerable to dissipation in
the absence of a regenerating dynamo.  By considering the evolution of a
small seed field in an era during which such a dynamo is not yet
functioning, or when its effect is negligible (i.e., the protogalactic
era), we develop a tighter bound; that is, we require a smaller upper
bound on the flux of
monopoles to avoid extinguishing the existing Galactic field.  
The time evolution
of the magnetic field in the protogalaxy is governed by competition
between amplification due to flux freezing, and dissipation due to a
possible flux of magnetic monopoles. The details of this competition may
be modeled by an equation of motion for the magnetic field of the form
\cite{adams}.

\begin{equation}
\frac{dB}{dt} = \gamma_{coll}B - \frac{g{\cal F}}{1+ \mu/B}
\end{equation}

Here the first term on the right hand side 
describes the field amplification due
to flux freezing and the second term describes dissipation
by a flux of monopoles.
Each of the quantities has been written in non-dimensional
units. Here $g$ is the magnetic charge in Dirac units
(we take $g=1$).  The
magnetic field is measured in units of the present day Galactic field
strength, \(3 \times 10^{-6} G \).  The parameter \(\gamma_{coll} \) 
represents the growth rate of the galactic field, and has units of
\(10^{-8}\mbox{ yr}^{-1} \)  ({\em i.e.} the Galactic rotation rate).  The
flux,
$\cal F$, is measured in units of $1.2 \times 10^{-16} \mbox{ cm}^{-2}
\mbox{sec}^{-1} \mbox{sr}^{-1}$. 
Finally, the dissipation term depends upon $\mu$, where $\mu = mv^2/\ell$;
$m$ is the monopole mass in units of $10^{17}$ GeV, $v$ is the monopole
velocity in units of $10^{-3}c$, and $\ell$ is the coherence length of the
Galactic field measured in units of 1 kpc \cite{adams,turner}.  

As the protogalaxy collapses, we expect that the only significant
amplification mechanism will be flux freezing.  This results in an
effective growth rate, \(\gamma_{coll} \approx 2/\tau_{coll} \), where
\(\tau_{coll} \) is the collapse time of the galaxy, approximately
$10^9$ yrs \cite{silk}.  In the dimensionless units just defined,
$\gamma_{coll} = 0.2$.  Dynamo action during this era is negligible
since, prior to collapse, the rotation period is extremely large
compared to the collapse time.  The rotation period today is
$10^8$yr; using conservation of angular momentum, we see that
during the protogalactic epoch, when the radius was larger by
a factor of roughly 50, the rotation period was $>10^{11}$yr.
In comparison, as mentioned above, the collapse timescale is roughly $10^9$yr.
Consequently, our results are independent
of any specific dynamo model.  Additional dissipative effects which
depend upon the magnitude of the Galactic magnetic field, such as
turbulent dissipation (which evolves as $B^2$), may also be ignored. 

One point of concern is the value of the coherence length of the Galactic
field during the protogalactic era, as the dissipative term in equation
(1) depends upon $\mu = mv^2/\ell g$.  Before Galactic collapse, we expect
this coherence length to be much longer than today's value of $\sim 1$
kpc.  Although the value is highly uncertain, we expect $l \sim 30-100$ kpc;
hence, throughout the paper we adopt the value $\ell =50$ kpc.

The monopole velocity $v$ is in units of $10^{-3}$c, as mentioned above.
We expect a massive monopole to acquire this velocity due
to gravitational acceleration by the Galaxy during infall.
In a model of galaxy formation in which the dark matter
aggregates first into a dark halo, and the baryons subsequently
fall into the potential well provided by the dark matter,
it is reasonable to assume that the monopoles attained
the virial velocity of $10^{-3}$c once the dark matter haloes
came into existence. Subequently, the protogalactic collapse
of the baryons drags the magnetic field with it. 
If some other model of galaxy formation is considered
in which the dark haloes do not form first, then it is possible
that the monopoles were moving somewhat slower; a slower
monopole velocity will only lead to a tighter bound on the
monopole flux. Hence we will use the value $10^{-3}$ as the
lowest monopole velocity, in order to be on
the conservative side.

Light monopoles ($\mu << B_0$) were accelerated to higher
velocities by the galactic field, while `heavy' monopoles
($\mu >> B_0$) did not have their velocities changed significantly.
For our fiducial values of $B_0 \sim 10^{-9}$G and $\ell =50$kpc, 
we expect that monopoles heavier than $2 \times 10^{15}$ GeV were moving at
$10^{-3}$ while lighter monopoles were moving faster.
As seen in figure 1 and discussed further below, 
the monopole flux bound changes slope at this value of the mass:
the bound is linear as a function of mass for monopoles heavier than
$2 \times 10^{15}$ GeV and flat for lower masses.

Given equation (2), the possible behavior of the field is straightforward
to determine.  If the field is to survive, we must have $dB/dt >0$.  This
requires
\begin{equation}
{\gamma_{coll}B + (\gamma_{coll} \mu - g {\cal F})>0}
\end{equation}
If the monopole flux ${\cal F} < \mu \gamma_{coll}$, then the
field survives for all initial values of the field strength.
This behavior holds for $\mu > B_0$, i.e., for heavy monopoles
with $m > 2 \times 10^{15} \bigl({B_0 \over 10^{-9}{\rm G}}\bigr)$ GeV ,
and gives rise to the linear dependence of the monopole flux bound on the mass
(see the Figure).
If, on the other hand, ${\cal F} > \mu \gamma_{coll}$, then
only initial field strengths $B_0 > B_c \equiv {\cal F}/\gamma_{coll} - \mu$
will survive.  This behavior holds for $\mu < B_0$, i.e., for
light monopoles with $m < 2 \times 10^{15}
\bigl({B_0 \over 10^{-9}{\rm G}}\bigr)$ GeV,
and gives rise to the flat part of the
monopole flux bound as a function of mass (see the Figure).
Thus, for the protogalactic field $B_0$ to survive during
collapse, the
flux of monopoles {\em at this time} must obey one of two bounds

\begin{equation}
{{\cal F} < \mu \gamma_{coll} \qquad\mbox{or}}
\end{equation}

\begin{equation}
{{\cal F} < (B_0 + \mu)\gamma_{coll}} \, .
\end{equation}

\noindent We note that the flux of monopoles is ${\cal F} = nv/(4\pi)$, where
$n$ is the monopole number density, which scales with the redshift as $n
\sim (1+z)^3$.  The flux today is smaller than 
the flux at the time of protogalactic collapse 
by a factor of $(1+z_{proto})^3 \approx 100$, where $z_{proto}$ is
the redshift of galaxy formation.  Hence the flux of monopoles
today is constrained to be smaller than the bounds of eqns. (4) and (5)
by this factor.

Thus we obtain an analytical estimate for the bound on monopole flux,

\begin{equation} {{\cal F} < 5 \times 10^{-21} \left( \frac{m}{10^{17}
\mbox{GeV}} \right) \mbox{cm}^{-2} \mbox{sec}^{-1} \mbox{sr}^{-1}}
\end{equation}
for $m > 2 \times 10^{15}$GeV, and 

\begin{equation}
{{\cal F} < 9 \times 10^{-23} \left(\frac{B_{0}}{10^{-9}\mbox{G}}
\right) \mbox{cm}^{-2} \mbox{sec}^{-1}
\mbox{sr}^{-1}}
\end{equation}

\noindent for $m < 2 \times 10^{15}$GeV.

\section{Discussion and Conclusion}

Without the support of a regenerative  mechanism, the Galactic
magnetic field is particularly susceptible to dissipation by a flux of
magnetic monopoles.  We have exploited this vulnerability of
a small B field early in the history of the Galaxy
to arrive at strict bounds on the monopole flux given
in eqns. (6) and (7).
Figure 1 summarizes the new flux bounds,  displays previous
Parker bounds, and has a line indicating a closure density of monopoles.
Here, $\Omega_m = \rho_m/\rho_c$ is the mass density of monopoles 
$\rho_m$ in units of the critical density 
required to close the universe, $\rho_c = 2.8 \times 10^{-29}$ g cm$^{-3}$ 
(for Hubble constant $H_o = 70$ km s$^{-1}$ Mpc$^{-1}$).

Note that the new protogalactic extension of the
Parker bound differs from the extended Parker bound 
even at low monopole masses, where the curves flatten out.
The reason for this discrepancy is twofold: i) the collapse
timescale in eqn. (2) is $1/\gamma_{coll} \sim 10^9$ yr,
whereas the corresponding rotation timescale in the extended
Parker bound is 200Myr; ii) we have taken the coherence
length of the magnetic field in the protogalaxy to be
50 kpc, whereas the coherence length in the Galaxy is
roughly 1 kpc.


As a caveat, note that the bounds in this paper rest
on the assumption that no amplifying
mechanism other than flux freezing exists during collapse.  Recent
work \cite{kulsrud} suggests that a preglactic seed field, amplified by a
post-collapse dynamo may not be the only way to generate the observed field.
Kulsrud {\em et al.} suggest the possibility that Kolmogorov turbulence
and flux freezing together, acting on a small seed field can produce the
observed Galactic field, without recourse to a large scale dynamo.  In
this case, the equation of motion becomes

\begin{equation}
\frac{dB}{dt} = (\gamma_{coll}+ \gamma_{turb})B - \frac{g{\cal F}}{1+
\mu/B}
\end{equation}

\noindent where the $\gamma_{turb}$ factor measures field growth due to
Kolmogorov turbulence, and $\gamma_{turb} \sim 600 (\rho_B/\rho_D)$ 
in the previously defined
units; here, $\rho_B/\rho_D$ is the ratio of baryonic matter to dark matter.
Hence, looking at eqns (4) and (5) one might
conclude that in this scenario the flux bounds would be weaker
by at least this factor.  However, there is another more important difference
between the scenario of Kulsrud {\it{et al.}} and the flux freezing
we have described above: in the Kulsrud {\it{et al.}} model,
the coherence length of the magnetic field during
this era is much smaller, on the order of viscous length scales.  If 
indeed $\ell$ is this small, then eqns. (5) and (6) do not result
in an interesting bound.  Monopoles would only begin to dominate
the contribution to $dB/dt$ once $\ell$ becomes large; currently,
in this model, it is unclear when this would happen.  However,
numerical simulations employed in \cite{kulsrud} lack the
resolution to demonstrate that homogenous Kolmogorov turbulence actually
occurs down to the smallest scales, so the final word on this issue must
wait until alternate theoretical methods are developed or computational
power is increased.

Modulo this caveat, we have found a protogalactic extension to the
Parker bound that is four and a half orders of magnitude
more stringent than the extended Parker bound \cite{adams}
for a given monopole mass.

Monopoles with $\Omega_m < 1$ that satisfy the extended
Parker bound of Ref. \cite{adams} were in principle accessible
to direct searches (that rely on electromagnetic interactions
of monopoles) in an experimental apparatus like MACRO
with longer running time.  However, monopoles with $\Omega_m <1$
that satisfy the new protogalactic extension of the Parker
bound presented in this paper are no longer accessible to
existing direct search experiments.  
The new maximum monopole flux at the closure bound is
${\cal F} = 8 \times 10^{-19}$ cm$^{-2}$ s$^{-1}$ sr$^{-1}$
and occurs for monopole mass near the Planck scale $\sim 10^{19}$GeV.
This maximum flux at the closure bound
is a factor of 150 lower than the maximum 
allowed by the extended Parker bound, which
occurs for a monopole mass near the GUT scale $\sim 10^{17}$GeV. Thus
MACRO would have to be scaled up in size by a factor of 
more than 100 to access the new maximum flux compatible
with the protogalactic extension as well as
the closure bound.  Hence the new monopole bound in this paper, while
more uncertain than previous bounds, would have serious
consequences for experimental searches.

\bigskip
\centerline{\bf Figure Caption}
Monopole flux limits as a function of the monopole mass $m$
in GeV.  The line labeled TPB Bound shows the modified Parker
bound obtained in Ref. \cite{turner}.  The dotted lines show
the extended Parker bound of Ref. \cite{adams}.  The solid lines 
show the protogalactic extension of the Parker bound of this
paper.  The line labeled $\Omega_M$ represents the bound obtained
by assuming monopoles are uniformly distributed throughout the
Universe but do not `over close' the Universe.  Here, we have
taken $H_o = 70$ km s$^{-1}$ Mpc$^{-1}$. If the monopoles
are clustered with galaxies, this closure bound becomes weaker by
a factor of $10^5$.

\bigskip
\centerline{\bf Acknowledgements}
We would like to thank the Department of Energy 
for support at the University of Michigan. K. Freese would
like to thank the Max Planck Institut fuer Physik in Muenchen,
where some of this work was completed, for hospitality
during her stay.  We also thank F. Adams, G. Laughlin, and M. Turner
for useful conversations.

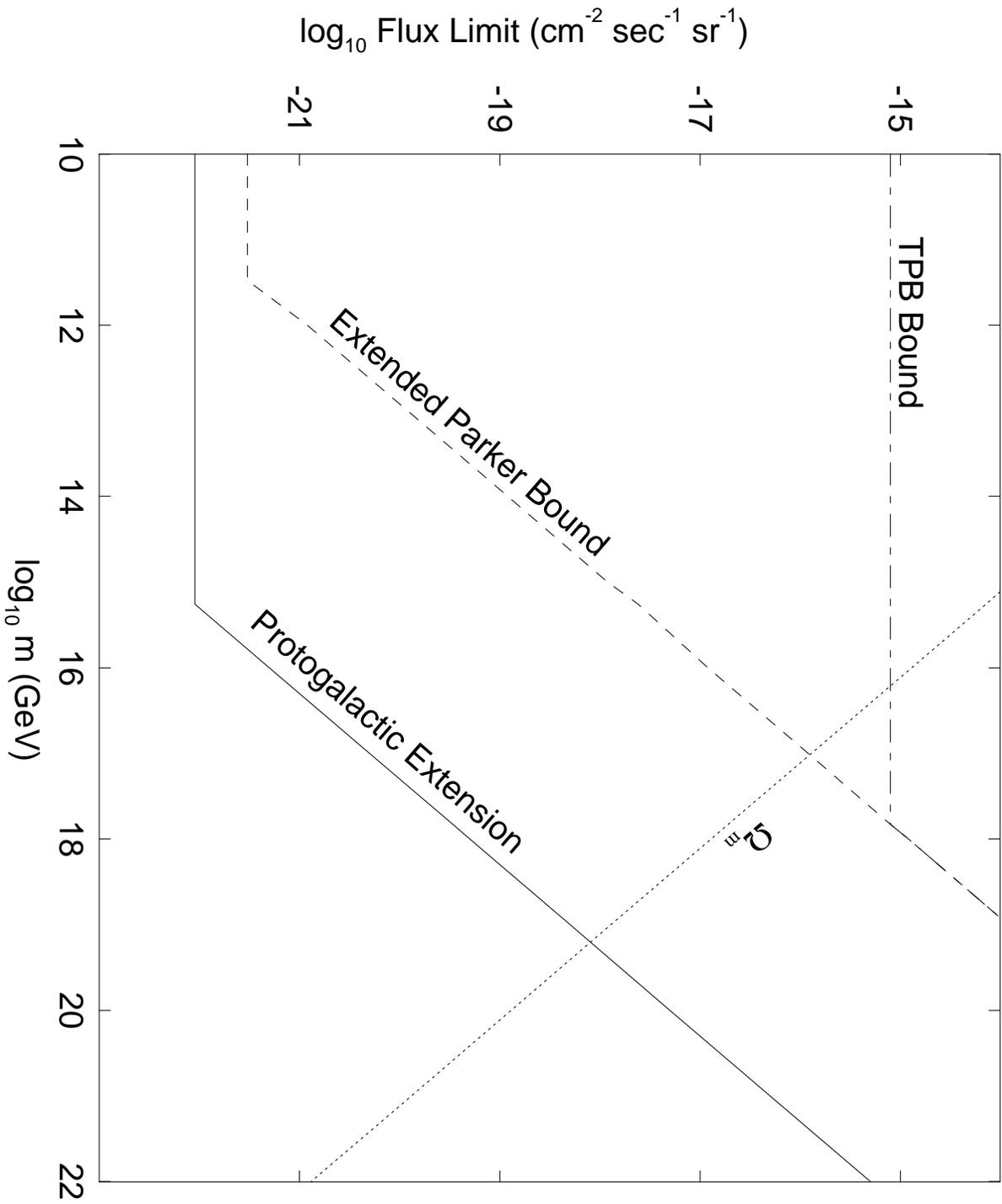
\end{document}